\documentclass[11pt, oneside]{amsart}   	
\usepackage{geometry}                		
\usepackage[parfill]{parskip}    		
\usepackage{graphicx}				
\usepackage{amssymb}
\usepackage[sort,comma,super,numbers]{natbib}
\usepackage{url}

\usepackage{multirow,array}
\usepackage{lscape} 

\usepackage{xcolor}

\title[Multiple imputation with raking of weights]{Combining multiple imputation with raking of weights: An efficient and robust approach in the setting of nearly-true models}

\author{Kyunghee Han}
\address{Department of Biostatistics, Epidemiology, and Informatics, University of Pennsylvania Perelman School of Medicine}
\email{kyunghee.stat@gmail.com}

\author{Pamela A. Shaw}
\address{Department of Biostatistics, Epidemiology, and Informatics, University of Pennsylvania Perelman School of Medicine}
\email{shawp@upenn.edu}

\author{Thomas Lumley}
\address{Department of Statistics, University of Auckland}
\email{t.lumley@auckland.ac.nz}
\thanks{Corresponding author: Kyunghee Han (\url{kyunghee.stat@gmail.com})}

\graphicspath{{figures/}}

\begin{document}						

\maketitle

\begin{abstract}
Multiple imputation provides us with efficient estimators in model-based methods for handling missing data under the true model. It is also well-understood that design-based estimators are robust methods that do not require accurately modeling the missing data; however, they can be inefficient. In any applied setting, it is difficult to know whether a missing data model may be good enough to win the bias-efficiency trade-off. Raking of weights is one approach that relies on constructing an auxiliary variable from data observed on the full cohort, which is then used to adjust the weights for the usual Horvitz-Thompson estimator. Computing the optimally efficient raking estimator requires evaluating the expectation of the efficient score given the full cohort data, which is generally infeasible. We demonstrate multiple imputation (MI) as a practical method to compute a raking estimator that will be optimal. We compare this estimator to common parametric and semi-parametric estimators, including standard multiple imputation. We show that while estimators, such as the semi-parametric maximum likelihood and MI estimator, obtain optimal performance under the true model, the proposed raking estimator utilizing MI maintains a better robustness-efficiency trade-off even under mild model misspecification. We also show that the standard raking estimator, without MI, is often competitive with the optimal raking estimator. We demonstrate these properties through several numerical examples and provide a theoretical discussion of conditions for asymptotically superior relative efficiency of the proposed raking estimator.
\end{abstract}

\section{Background}
In many settings, variables of interest maybe too expensive or too impractical to measure precisely on a large cohort. Generalized raking is an important technique for using whole population or full cohort information in the analysis of a subsample with complete data, \citep{deville1992calibration, sarndal2007calibration, breslow2009using} closely related to the augmented inverse probability weighted (AIPW) estimators of Robins and co-workers.\citep{robins1994estimation, firth1998robust, lumley2011connections} Raking estimators use auxiliary data measured on the full cohort to adjust the weights of the Horvitz-Thompsonn estimator in a manner that leverages the information in the auxiliary data and improves efficiency. The technique is also, and perhaps more commonly, known as ``calibration of weights'', but we will avoid that term here because of the potential confusion with other uses of the word ``calibration''. An obvious competitor to raking is multiple imputation of the non-sampled data.\citep{rubin1996multiple}  While multiple imputation was initially used for relatively small amounts of data missing by happenstance, it has more recently been proposed and used for large amounts of data missing by design, such as when certain variables are only measured on a subsample taken from a cohort.\citep{marti2011multiple, keogh2013using, jung2016fitting, seaman2012combining, morris2014tuning}

In this paper we take a different approach. We use multiple imputation to construct new raking estimators that are more efficient than the simple adjustment of the sampling weights \cite{breslow2009using} and compare these estimators to direct use of multiple imputation in a setting where the imputation model may be only mildly misspecified.  Our work has connections to the previous literature, where multiple imputation and empirical likelihood are used in the missing data paradigm to construct multiply robust estimators that are consistent if any of a set of imputation models or a set of sampling models are correctly specified.\cite{han2016combining} We differ from this  work in assuming known subsampling probabilities, which allows for a complex sampling design from the full cohort, and in evaluating robustness and efficiency under contiguous (local) misspecification following the ``nearly-true models'' paradigm.\cite{lumley2017robustness} Known sampling weights commonly arise in settings, such as retrospective cohort studies using electronic health records (EHR) data, where a validation subset is often constructed to estimate the error structure in variables derived using automated algorithms rather than directly observed. Lumley (2017) \cite{lumley2017robustness} considered the robustness and efficiency trade-off of design-based estimators versus maximum likelihood estimators in the setting of nearly-true models. We build on this work by comparing multiple imputation with the standard raking estimator, and examine to what extent raking that makes use of multiple imputation to construct the auxiliary variable may affect the bias-efficiency trade-off for this setting. 

We first introduce the raking framework in Section 2. In Section 3, we describe the proposed raking estimator, which makes use of multiple imputation to construct the potentially optimal raking variable. In Section 4, we compare design-based estimators with standard multiple imputation estimators in two examples using simulation, a classic case-control study and a two phase study where the linear regression model is of interest and an erroprone surrogate is observed on the full cohort in place of the target variable.  For this example, we additional study the relative performance of regression calibration, a popular method to address covariate measurement error. \citep{carroll2006} In section 5, we consider the relative performance of multiple imputation versus raking estimators in the National Wilms Tumor Study. We conclude with a discussion of the robustness efficiency trade-off in the studied settings.

\section{Introduction to raking framework}
Assume a full cohort of size $N$ and a probability subsample of size $n$ with known sampling probability $\pi_i$ for the $i$-th individual. Further, assume we observe an outcome variable $Y$, predictors $Z$, and auxiliary variables $A$ on the whole cohort, and observe predictors $X$ only on the sample. Our goal is to fit a model $P_\theta$ for the distribution of $Y$ given $Z$ and $X$ (but not $A$).  Define the indicator variable for being sampled as $R_i$. We assume an asymptotic setting in which as $n\to\infty$, a law of large numbers and central limit theorem exist.  In some places we will make the stronger asymptotic assumption that the sequence of cohorts are iid samples from some probability distribution and that the subsamples satisfy $\inf_i \pi_i>0$.\cite{breslow2009using,lumley2011connections,lumley2017robustness} 

With full cohort data with complete observations we would solve an estimating equation
\begin{equation}
\sum_{i=1}^N U(Y_i,X_i,Z_i;\theta)=0,
\label{eq-census}
\end{equation}
where $U = U(Y,X,Z;\theta)$ is an estimate of the efficient score or influence function  for giving at least locally efficient estimation of $\theta$ with complete data. We write $\tilde\theta_N$ for the resulting estimator with complete data from the full cohort, and assume it converges in probability to some limit $\theta^*$. If the cohort is truly a realization of the model $P_\theta$ we write $\theta_0$ for the true value of $\theta$. We assume $\tilde\theta_N$ would be a locally efficient estimator in the model $P_\theta$ at $\theta_0$, given compete data. 

The Horvitz-Thompson-type estimator  $\hat\theta_{HT}$ of $\theta$ solves
\begin{equation}
\sum_{i=1}^N \frac{R_i}{\pi_i}U(Y_i,X_i,Z_i;\theta)=0.
\label{eq-ht}
\end{equation}
Under regularity conditions, for example the existence of a central limit theorem and sufficient smoothness for $U$, it is also consistent for  $\theta^*$, and thus for $\theta_0$ if $P_\theta$ is correctly specified. 

A generalized raking estimator using an auxiliary variable $H=H(Y, Z,A;\eta)$, which may depend on some parameter $\eta$, solves a weighted estimating equation
\begin{equation}
\sum_{i=1}^N \frac{g_iR_i}{\pi_i} U(Y_i,X_i,Z_i;\theta)=0,
\label{eq-aipw}
\end{equation}
where the weight adjustments $g_i$ are chosen to satisfy the calibration constraints
\begin{align}
\sum_{i=1}^N \frac{R_ig_i}{\pi_i} H(Y_i, Z_i,A_i;\eta) = \sum_{i=1}^N H(Y_i, Z_i,A_i;\eta) \label{cal-adj}
\end{align}
while minimizing a distance function $\sum_{i=1}^n d(g_i/\pi_i, 1/\pi_i)$. Lagrange multipliers can be used to construct an iteratively weighted least squares algorithm for computing $g_i$.\cite{deville1992calibration} 

In the standard multiple imputation, we use a model for the distribution of $X$ given $Z$, $Y$ and $A$. For this, we generate $M$ samples from the predictive distribution to produce $M$ imputations $X_i^{(1)},\ldots, X_i^{(M)}$, giving rise to $M$ complete imputed datasets that represent samples from the unknown conditional distribution of the complete data given the observed data.  It is now straightforward to solve equation (\ref{eq-census}) for each of the $m$-th imputed dataset, giving $M$ values of $\tilde{\theta}_{N,(m)}$ with estimated variances $\tilde{\sigma}_{N,(m)}^2$, $1 \leq m \leq M$.  The imputation estimator $\hat\theta_{\mathrm{MI}}$ of $\theta$ is the average of the $\tilde{\theta}_{N,(m)}$, and the variance can be estimated from the variance of the $\tilde{\theta}_{N,(m)}$ and the average of $\tilde{\sigma}_{N,(m)}^2$.\citep{rubin1996multiple}

\section{Imputation for calibration} \label{impute-cal}

\subsection{Estimation}

The optimal function $H_i$ is $E[U_i|Y_i, Z_i, A_i]$, and using this optimal $H_i$ would give the optimal design-consistent estimator of $\theta$,\citep{robins1994estimation}. However, the optimal $H_i$ is typically not available explicitly. In practice, one may estimate the optimal function $H_i$ with a single regression imputation $\hat X_i$ of $X_i$, where we first solve 
$$\sum_{i=1}^N U(Y_i,\hat X_i,Z_i;\theta)=0,$$
with respect to $\theta$, and then, compute $U(Y_i,\hat X_i,Z_i;\theta)$ at the solution.\cite{breslow2009using,rivera2016using} We write such a calibration estimator of $\theta$ with a single regression imputation by $\hat\theta_{\mathrm{cal,1}}$.

In this study, we propose a raking estimator using multiple imputation. Specifically, we first solve the sets of equations
$$\sum_{i=1}^N U(Y_i,\hat{X}_i^{(m)},Z_i;\theta)=0,$$
where $\hat{X}_1^{(m)}, \ldots, \hat{X}_N^{(m)}$ are imputed values of $X_i$ for each $m$-th imputation procedure to get multiple estimates $\hat\theta^{(m)}$, $1 \leq m \leq M$. Define $H_i$, for each $1 \leq i \leq N$, as the average of the $M$ resulting $U(Y_i,\hat{X}_i^{(m)},Z_i;\hat\theta^{(m)})$:
\begin{align}
H_i = \frac{1}{M} \sum_{m=1}^M U(Y_i,\hat{X}_i^{(m)},Z_i;\hat\theta^{(m)}). \label{multical-adj}
\end{align}
Finally, we solve \eqref{eq-aipw} with the weight adjustments under the calibration constraint \eqref{cal-adj}, and write the final estimator $\hat{\theta}=\hat\theta_{\mathrm{cal,M}}$ of $\theta$.

\subsection{Efficiency and robustness} \label{efficient-robust}

When all three of the sampling probability,  the imputation model, and the regression model are correctly specified, the standard calibration estimator $\hat\theta_{\mathrm{cal,1}}$ gives a way to compute the efficient design-consistent estimator. If we are willing to only assume the regression model and imputation model are correct, there appears to be no motivation for requiring a design-consistent estimator. In this case, the standard multiple imputation estimator $\hat\theta_{\textrm{MI}}$ will also be consistent and typically more efficient than a design-based approach.  

If the regression model and the imputation model are correctly specified with all the available variables, it is clear that the empirical average \eqref{multical-adj} over multiple imputations in $H_i$ will converge to the optimal value $E[U_i|Y_i,Z_i, A_i]$ as $M$ and $N$ increase, so that the proposed raking estimator using multiple imputation provides the optimal calibration estimator. However, it is unreasonable in practice to assume that both the regression and imputation models are exactly correct. Recently, in the special case where the full cohort is an iid sample and the subsampling is independent, so-called Poisson sampling, it has been shown that the inverse probability weighting adjusted by multiple imputation attains the semi-parametric efficiency bound for a model that assumes only $E[U_i]=0$ and $E[R_i|Z_i,Y_i,A_i]=\pi_i$,\cite{han2016combining} where the proposed estimator $\hat\theta_{\mathrm{cal,M}}$ also solves a weighted estimating equation \eqref{eq-aipw} subject to the calibration constraints \eqref{cal-adj} computed by multiple imputation.

In this paper, we argue one step further that the interesting questions of robustness and efficiency arise when the imputation model and potentially also the regression model are slightly misspecified. Under what conditions are $\|\hat\theta_{\mathrm{cal,M}}-\theta^*\|_2^2$ and $\|\hat\theta_{\mathrm{MI}}-\theta^*\|_2^2$ comparable, and do these correspond to plausible misspecifications of the regression model, the imputation model, or both? These questions were considered in a more abstract context by Lumley (2017)\cite{lumley2017robustness}, where the model is only nearly-true such that
$$\sqrt{n}(\hat\theta_{\mathrm{cal,M}}-\theta^*){\rightsquigarrow} N(0,\sigma^2+\omega^2)$$
and
$$\sqrt{n}(\hat\theta_{\mathrm{MI}}-\theta^*){\rightsquigarrow} N(\kappa\rho\omega,\sigma^2).$$
In the above equations, $\kappa$ is the limit of Kullback--Leibler divergence between the true model $P_n$ and the outcome model $Q_n$ defined as the sequence of misspecified distributions chosen to be contiguous to the true model. We assume $\kappa$ is bounded. $\rho$ is the asymptotic correlation between the log-likelihood ratio of two distributions, $P_n$ and $Q_n$, and the difference in influence functions for $\hat\theta_{\mathrm{cal,M}}$ and $\hat\theta_{\mathrm{MI}}$ under $P_n$ and $Q_n$, respectively. That is, the ``nearly-true'' models are defined by a sequence of outcome models such that one may not reliably reject misspecification, even using the most powerful test comparing the truly data-generating distribution. In simple but common cases, including the case-control design study and the linear regression analysis in the two-phase study, the model misspecification may neutralize the advantage of the standard multiple imputation. \cite{lumley2017robustness} Indeed the mean-squared error of $\hat\theta_{\mathrm{MI}}$ will be asymptotically larger than that for $\hat\theta_{\mathrm{cal,M}}$ whenever $|\kappa \rho|>1$.\cite{lumley2017robustness}  We study the relative numerical performance of these two estimators and other standard competitors under nearly-true model setting in the next section.

\section{Simulations} \label{sec-sim}
In this section we are interested in three questions; how much precision is gained by multiple versus single imputation in raking, whether imputation models can maintain an efficiency advantage while being more robust, and how these affect the efficiency-robustness trade-off between weighted and imputation estimators. Source code in R for these simulations is available at \url{https://github.com/kyungheehan/calib-mi}.

\subsection{Case-control study}\label{sim1}
We first demonstrate numerical performance of multiple imputation for the case-control study where calibration is not available but the maximum likelihood estimator can be easily computed. Let $X$ be a standard normal random variable and $Y$ be a binary response taking values in $\{0,1\}$ such that for a given $X=x$ the associated logistic model is given by
\begin{align}
	\textrm{logit}\,\mathbb{P}(Y=1 | X=x) = \alpha_0 + \beta_0 x + \delta_0(x-\xi) \mathbb{I}(x > \xi) \label{true1}
\end{align}
for some fixed $\delta_0$ and $\xi$, and $\textrm{logit}(p) = \log \big( \frac{p}{1-p} \big)$ for $0 < p < 1$. In accordance with the usual case-control study design, we assume $Y$ is known for everyone, but $X$ is available with sampling probability of 1 when $Y=1$ and a lower sampling probability when $Y=0$. To be specific, we first generate a full cohort $\mathcal{X}_N = \{ (Y_i, X_i) :  1 \leq i \leq N \}$ following the true model \eqref{true1} and denote the index set of all the $n$-case subjects in $\mathcal{X}_N$ by $S_1 \subset \{ 1, \ldots, N \}$, $n < N$. Thus,  $Y_i=1$ if $i \in S_1$, otherwise $Y_i = 0$. Then a balanced case-control design is employed which consists of observing $(Y_i, X_i)$ for all the subjects in $S_1$ and a randomly chosen $n$-subsample $S_0$ from $\{1, \ldots, N \} \setminus S_1$. For cohort members $\{1, \ldots, N \} \setminus S_0 \cup S_1$,  only $Y_i$ is observed. Define $\mathcal{X}^\ast_n = \{ (Y_i, X_i) : i \in S_0 \cup S_1 \}$.

We examine the sensitivity of the multiple imputation approach in the setting of nearly-true models.\citep{lumley2017robustness} For a practical definition of a nearly-true model, we consider a working model that may not be reliably rejected, even when using the oracle test statistic of the likelihood ratio with the true model  \eqref{true1} used to generate the data as the null. In other words, instead of fitting the true model \eqref{true1}, we employ a simpler outcome model 
\begin{align}
	\textrm{logit} \, \mathbb{P}(Y=1 | X=x) = \alpha + \beta x.
\label{nearly-true1}
\end{align}
We note that when $\delta_0=0$ the working model \eqref{nearly-true1} is correctly specified, but misspecified when $\delta_0\neq 0$. It is worthwhile to mention that the single knot linear spline logistic model \eqref{true1} is the worst case of misspecified model of \eqref{nearly-true1} when $\alpha_0 = -5$, $\beta_0 = 1$ and $\xi \approx 1.8$, which maximizes correlation between the most powerful test to reject the model misspecification and the bias of the misspecified maximum likelihood estimator. \citep{lumley2017robustness} In this case, the maximum likelihood estimator of \eqref{nearly-true1} is the unweighted logistic regression \citep{prentice1979logistic} for the complete case analysis only with $\mathcal{X}_n^\ast$.

Four different methods are compared in our example for estimating the nearly-true slope $\beta$ in \eqref{nearly-true1}; (i) the maximum likelihood estimation (MLE), (ii) a design-based inverse probability weighting (IPW) approach, (iii) a multiple imputation with a parametric imputation model (MI-P) and (iv) a multiple imputation with non-parametric imputation based on  bootstrap resampling (MI-B). Formally, the parametric MI (MI-P) imputes covariates $X_i$, $i \not\in S_0\cup S_1$, from a parametric model such that $X|Y=y$ is assumed to be distributed as $N(\mu + \eta y, \sigma^2)$, where $\mu = \mathbb{E}(X | Y=0)$, $\eta = \mathbb{E}(X | Y=1) - \mu$, and $\sigma^2 = \mathbb{V}\text{ar}(X)$. Here, the parameters $\mu$, $\eta$ and $\sigma^2$ are estimated from $\mathcal{X}_n^\ast$. On the other hand, the bootstrap method (MI-B) resamples covariates $X_i$, $i \not\in S_0\cup S_1$, from the empirical distribution of $X$ given $Y=0$. We note that MLE only utilizes the sub-cohort information $\mathcal{X}_n^\ast$ but the other estimators additionally use response observations $\{Y_i : i \not\in S_0 \cup S_1\}$ so that efficiency gains can be expected for estimating the nearly-true slope $\beta$, depending on the level of model misspecification.

Using Monte Carlo iterations, we summarized the empirical performance of the four different estimators based on fitting the nearly-true model \eqref{nearly-true1} with the mean squared error (MSE) of the target parameter $\beta$,
\begin{align}
	\textrm{MSE}(\hat{\beta}) = \frac{1}{K} \sum_{k=1}^K \big( \hat{\beta}^{[k]} - \beta \big)^2 \label{mse},
\end{align}
where $\hat{\beta}^{[k]}$ is the estimate of $\beta$ from the $k$-th Monte Carlo replication, $1 \leq k \leq K$. Similarly the empirical bias-variance decomposition,
\begin{align}
	\textrm{Bias}(\hat{\beta}) = \textrm{E}{\hat{\beta}} - \beta \quad \textrm{and}\quad \textrm{Var}(\hat{\beta}) = \frac{1}{K} \sum_{k=1}^K \Big( \hat{\beta}^{[k]} - \textrm{E}{\hat{\beta}} \Big)^2, \label{bias-var}
\end{align}
was also reported to compare precision and efficiency, where $\textrm{E}{\hat{\beta}} = K^{-1} \sum_{k=1}^K \hat{\beta}^{[k]}$. For all simulations, we fixed $\beta=1$, $\alpha_0=-5$, $\xi_0=1.8$, $N=10^4$, and the number of cases was around $n=110$ in average. We used $M=100$ multiple imputations and $K=1000$ Monte Carlo simulations. Results are provided in Table \ref{table1}.

Table \ref{table1} demonstrates two principles. First, the parametric MI (MI-P) estimator closely matches the maximum likelihood estimator, but the resampling (MI-B) estimator closely matches the design-based estimator.  Second, more importantly, the design-based estimator is less efficient than the maximum likelihood estimator when the model is correctly specified, but has lower mean squared error when $\delta_0$ was greater than about $1.6$. In this case, even the most powerful one-sided test of the null $\delta_0=0$ based on the alternative model \eqref{nearly-true1} would have power less than approximately $0.5$, so that any model diagnostic used in a practical setting would have lower power. Figure 1 shows the relative efficiency of the methods as a function of the level of mispecification. In summary, we conclude that the efficiency gain of the model-based analysis is not robust even to mild forms of misspecification that would not be detectable in practical settings.

\subsection{Linear regression with continuous surrogate}\label{sim2}

We now evaluate the performance of the multiple imputation raking estimator in a two-phase sampling design. Let $Y$ be a continuous response associated with covariates $X=x$ and $Z=z$ such that 
\begin{align}
	\mathbb{E}(Y | X=x, Z=z)= \alpha_0 + \beta_0 x + \delta_0 x \cdot \mathbb{I}(|z| > \zeta_0), \label{true2}
\end{align}
for some fixed $\delta_0$ and $\zeta_0 = F_Z^{-1}(0.95)$, where $\mathbb{V}ar(Y|X,Z)=1$, $X$ is a standard normal random variable,  $Z$ is a continuous surrogate of $X$ and $F_Z^{-1}$ is the inverse cumulative distribution fuction for $Z$. Similarly to the simulation study in the previous section \ref{sim1}, instead of the true model \eqref{true2} which generally will not be known in a real data setting, we are interested in the typical linear regression analysis with an outcome model 
\begin{align}
	\mathbb{E}(Y | X=x) = \alpha + \beta x. \label{nearly-true2}
\end{align}
Two different scenarios of the surrogate variable $Z$ are considered such that (a) $Z = X + \varepsilon$ for $\varepsilon \sim N(0,1)$ and (b) $Z= \eta X$ for $\eta \sim \Gamma(4,4)$, which represent additive and multiplicative error, respectively. In the first phase of sampling, we assume that outcomes $Y$ and auxiliary variables $Z$ are known for everyone, whereas covariate measurements of $X$ are available only at the second stage. The sampling for the second phase will be stratified on $Z$. Specifically, we will observe $X_i$ for all individuals if $|Z_i| > \zeta_0$,  otherwise $5\%$ of subjects subjects in the intermediate stratum $|Z_i| \leq \zeta_0$ are randomly sampled, where $1 \leq i \leq N$. We write $S_2 \subset \{1, \ldots, N \}$ to be the index set of subjects collected in the second phase so that $\mathcal{X}_I = \{ (Y_i, Z_i) :  1 \leq i \leq N \}$ and $\mathcal{X}_{II} = \{ (Y_i, X_i, Z_i) :  i \in S_2 \}$ denote the first and second stage samples, respectively. 

We compare five different methods of estimating the nearly-true parameter $\beta$: (i) maximum likelihood estimation (MLE), (ii) a standard generalized raking estimation using the auxiliary variable, (iii) regression calibration (RC), a single imputation method that imputes the missing covariate $X$ with an estimate of $\mathbb{E}[X|Z]$,\citep{carroll2006} (iv) multiple imputation without raking (MI), and (v) the proposed approach combining raking and the multiple imputation (MIR). We note that when $Y$ is Gaussian, the semi-parametric efficient maximum likelihood estimator of $\beta$ is available in the \texttt{missreg3} package in R,\citep{wild2013missreg3} using the stratification information.\cite{scott2006calculating} We employ this for the MLE (i).

For the standard raking method (ii), we construct a design-based efficient estimator \citep{breslow2009using} as below:
\begin{itemize}
	\item[R1.] Find a single imputation model $X = a + b Y + c Z + \epsilon$, where $\epsilon \sim N(0,\tau^2)$ based on the second phase sample $\mathcal{X}_{II}$.
	\item[R2.] Fit the nearly-true model \eqref{nearly-true2} using $(Y_i, \hat{X}_i)$ for $1 \leq i \leq N$, where $\hat{X}_i$ are fully imputed from (R1).
	\item[R3.] Calibrate sampling weights for raking using the influence function induced from the nearly-true fits in (R2).
	\item[R4.] Fit the design-based estimator of the nearly-true model \eqref{nearly-true2} with the second phase sample $\mathcal{X}_{II}$ and calibrated sampling weights from (R3).
\end{itemize}
For the conventional regression calibration approach (iii), we simply fit a linear model regressing $X_i$ on $Z_i$ for $i \in S_i$ and then impute missing observations $\hat{X}_i$ in the first phase so that the nearly-true model \eqref{nearly-true2} is evaluated using $\{ (Y_i, \hat{X}_i) :  i \not\in S_2\}$ and $\{ (Y_i, X_i) : i \in S_2 \}$.

We consider two resampling techniques for the multiple imputation method (iv): the wild bootstrap \citep{cao1991rate, mammen1993bootstrap,hardle1993comparing} and a Bayesian approach with a non-informative prior. Note, the wild bootstrap gives consistent estimates for settings where the conventional Efron's bootstrap does not work, such as  under heteroscedasticity and high-dimensional settings. We refer to Appendix \ref{App-cal} for implementation details of multiple imputation with the wild bootstrap and a parametric Bayesian resampling. We now illustrate the proposed method that calibrates sampling weights using multiple imputation. 
\begin{itemize}
	\item[M1.] Resample $\hat{X}_i^\ast$ independently for all $1 \leq i \leq N$ by using either the wild bootstrap or the parametric Bayesian resampling.
	\item[M2.] Fit the nearly-true model \eqref{nearly-true2} based on a resample $\{ (Y_i, \hat{X}_i^\ast) :  1 \leq i \leq N\}$.
	\item[M3.] Repeat (M1) and (M2) in multiple times, and take the average of influence functions, induced by the nearly-true models fitted in (M2).
	\item[M4.] Calibrate sampling weights using the average influence function as auxiliary information.
	\item[M5.] Fit the design-based estimator of the nearly-true model \eqref{nearly-true2} with the second phase sample $\mathcal{X}_{II}$ and calibrated sampling weights obtained from (M4).
\end{itemize}

Setting {$N=5000$}, we ran $M=100$ multiple imputations over {$1000$} Monte Carlo replications. For all simulations, $\beta=1$, $\alpha_0=0$, $\zeta_0\approx2.3$ when $Z$ is a surrogate of $X$ with an additive measurement error but $\zeta_0\approx1.8$ with a multiplicative error in our simulation settings, and the phase two sample with $|S_2|=750$ in average. We considered several values of $\delta_0$ and the level of misspecification is described by the empirical power to reject the misspecified model for the level $0.05$ likelihood ratio test  comparing the null \eqref{true2} and alternative \eqref{nearly-true2}.

 The numerical results with additive measurement errors are summarized in Table \ref{table2} and Figure \ref{figure2}. In this scenario, regression calibration (RC)  performed the best for $\delta_0$ less than approximately 0.15, since RC correctly assumes a linear model for imputing $X$ from $Z$. The two standard multiple imputation had estimation bias due to a misspecified imputation model and had a larger MSE than the RC method. However, we note once again the model diagnostic for linearity, i.e. $\delta_0=0$, had at most $20\%$ power for the level of misspecifictation studied, which means one may not reliably reject the misspecified model even when $\delta_0=0.3$ and imputation with the correctly specified model is also unlikely. Indeed the standard and proposed MIR raking estimators achieved lower MSE when $\delta_0 \geq 0.15$. Thus, raking successfully leveraged the information from the cohort not in the phase two sample while maintaining its robustness, as seen in previous literature.\citep{deville1992calibration, sarndal2007calibration, breslow2009using} In this simulation we further found that the standard raking estimation efficiency can be improved by using  multiple imputation to estimate the optimal raking variable, with efficiency gains of about $10\%$ in this example.  Table \ref{table3} and Figure \ref{figure3} summarize the results for the multiplicative error scenario. In this case, even for $\delta_0=0$, the RC and multiple imputations have appreciable bias and worse relative performance compared to the two raking estimators, because of the misspecified imputation model. The two raking estimators outperformed all estimators for all levels of misspecfication. In this scenario, the MIR had smaller gains over the standard raking estimator.

\section{Data Example: The National Wilms Tumor Study} \label{sec-data}

We apply our proposed approach to the data from  National Wilms Tumor Study (NWTS). In this example, we assume a key covariate of interest is only available in a phase 2 subsample, and compare the proposed MIR method with other standard estimators for this setting.   In the data example with NWTS, we are interested in the logistic model for the binary relapse response with predictors histology (UH: unfavorable versus FH: favorable versus), the stage of disease (III/IV versus I/II), age at diagnosis (year) and the diameter of tumor (cm) as 
\begin{eqnarray}
\begin{split}
	\qquad
	&\textrm{logit} \, \mathbb{P}(\textrm{Relapse} \, | \, \textrm{Histology}, \textrm{Stage}, \textrm{Age}, \textrm{Diameter})\\
	&\quad = \alpha + \beta_1 (\textrm{Age}) + \beta_2 (\textrm{Diameter}) + \beta_3 (\textrm{Histology}) + \beta_4 (\textrm{Stage}) + \beta_{3,4} (\textrm{Histology}\ast\textrm{Stage}),
\end{split} \label{wilms-model}
\end{eqnarray}
where $\beta_{3,4} $ indicates an interaction coefficient between histology and stage.\cite{lumley2011complex} We consider \eqref{wilms-model} is a nearly-true model of the relapse probability associated with covariates, as it is difficult to specify the true model in this real data setting.

Histology was evaluated from both a central laboratory and a local laboratory, where the latter is subject to misclassification due to the difficulty of diagnosing this rare disease. For the first phase data, we suppose that the $N=3915$ observations of outcomes and covariates are available for the full cohort, except that the histology is obtained only from the local laboratory. Central histology is then obtained on a phase 2 subset. By considering the outcome-dependent sampling strategies,\cite{breslow1999design,lumley2011complex} we sampled individuals for the second phase by stratifying on relapse, local histology and disease stage levels. Specifically, all the subjects who either relapsed or had unfavorable local histology were selected, while only a random subset in the remaining strata (non-relapsed and favorable histology strata for each stage level) were selected so that there was a  1:1 case-control sample for each stage level.\cite{lumley2011complex} 

Similarly to previous numerical studies, we compared four estimators, where the ``true parameters'' in \eqref{wilms-model} are given by estimates from the full cohort analysis: (i) the maximum likelihood estimates (MLE) of the regression coefficients in \eqref{wilms-model} based on the complete case analysis of the second phase sample; (ii) the standard raking estimator, which calibrates sampling weights by using the local histology information in the first phase sample, where the raking variable was generated by the influence functions. We imputed (unobserved) a central histology path by using a logistic model regressing the second phase histology observations on the age, tumor diameter and three-way interaction among the relapse, stage and local histology together with their nested interaction terms. The reason for introducing interaction in the imputation model is that subjects at advanced disease stage or with unfavorable histology were mostly relapsed in the observed data.  We also consider (iii) the conventional bootstrap procedure was employed for multiple imputation (MI) with the second phase sample, and (iv) we  combined the raking and multiple imputation (MIR) as proposed in the previous section. 

 The relative performance of the methods were assessed by obtaining estimates for 1000 two-phase samples. 100 multiple imputations were applied for each two-phase sample. Table \ref{table4} summarizes the results. Similarly to the numerical illustration in the previous section, we found that the proposed method (MIR) had the best performance in terms of achieving lowest MSE for the target parameter available only on the subset. While raking does not provide the lowest MSE for all parameters, in this example, MIR had the lowest squared error summed over the model parameters.

\section{Discussion}

There are many settings in which variables of interest are not directly observed, either because they are too expensive or difficult to measure directly or because they come from a convenient data source, such as EHR, not originally collected to support the research question. In any practical setting, the chosen statistical model to handle the mismeasured or missing data will be at best a close approximation to the targeted true underlying relationship. A general discussion of the difficulty of testing for model misspecification demonstrates that the data at hand cannot be used to reliably test whether or not the basic assumptions in the regression analysis hold without good knowledge of the potential structure.\cite{freedman2009} Here, we have considered the robustness-efficiency trade of several estimators in the setting of mild model misspecification, where idealized tests with the correct alternative have low power. When the misspecification is along the least-favorable direction contiguous to the true model, the bias will be in proportion to the efficiency gain from a parametric model.\cite{lumley2017robustness} We studied the relative performance of design-based estimators for a nearly-true regression model in two cases, logistic regression in a case-control study and linear regression in a two-phase design, where the misspecification was approximately in the least favorable direction. In both cases, the misspecification took the form of a mild departure from linearity, and as expected, the raking estimators demonstrated better robustness compared to the parametric MLE and standard multiple imputation models. 

Our approach to local robustness is related to that of Watson and Holmes (2016),\cite{watson2016} who consider making a statistical decision robust to model misspecification around the neighborhood of a given model in the sense of Kullback--Leibler divergence. Our approach is simpler than theirs for two reasons: we consider only asymptotic local minimax behavior, and we work in a two-phase sampling setting where the sampling probabilities are under the investigator's control and so can be assumed known.  In this setting, the optimal raking estimator is consistent and efficient in the sampling model and so is locally asymptotically minimax. In more general settings of non-response and measurement error, it is substantially harder to find estimators that are local minimax, even asymptotically, and more theoretical work is needed.

Another contribution of our study is that we demonstrated a practical approach for the efficient design-based estimator under contiguous misspecification. Without an explicit form of an efficient influence function, the characterization of the efficient estimator may not always lead to readily attainable computation of the efficient estimator in the standard raking method. We examined the use of multiple imputation to estimate the raking variable that confers the optimal efficiency. \citep{han2016combining} Our proposed raking estimator is easy to calculate and provides better efficiency than any raking estimator based on a single imputation auxiliary variable. In the two cases studied, the improvement in efficiency was evident, though at times small. On the other hand, the degree of improvement of the MI-raking estimator over the standard raking approach is expected to increase with the degree of non-linearity of the score for the target variable. In additional simulations, not shown, we did indeed see larger efficiency gains for MI-raking over single-imputation raking with large measurement error in $Z$.  

In many settings, there is a preference to choose simpler models when there is a lack of evidence to support a more complicated approach, because of the clarity of interpretation with simpler models. \citep{box2005statistics, stone1985additive} In such settings, design-based estimators are easy to implement in standard software and provide a desired robustness.  More theoretical work is also needed to find a more practical representation of the least-favorable contiguous model for the general setting in order to better understand how much of a practical concern this type of misspecification may be. The bias--efficiency trade-off we describe is also important in the design of two-phase samples.  The optimal design for the raking estimator will be different from the optimal design for the efficient likelihood estimator, and the optimal design when the outcome model is ``nearly-true'' may be different again.


\section*{Acknowledgments}
This work was supported in part by the Patient Centered Outcomes Research Institute (PCORI) Award R-1609-36207 and U.S. National Institutes of Health (NIH) grant R01-AI131771. The statements in this manuscript are solely the responsibility of the authors and do not necessarily represent the views of PCORI or NIH.

\section*{Data availability}
Source code in R for these simulations and the National Wilms Tumor Study data are available at \url{https://github.com/kyungheehan/calib-mi}.

%
%
%
%
%
%
%
%

\nocite{*}
\bibliographystyle{unsrtnat}
\bibliography{ref-multi-calib}%

\begin{thebibliography}{33}
\providecommand{\natexlab}[1]{#1}
\providecommand{\url}[1]{\texttt{#1}}
\expandafter\ifx\csname urlstyle\endcsname\relax
  \providecommand{\doi}[1]{doi: #1}\else
  \providecommand{\doi}{doi: \begingroup \urlstyle{rm}\Url}\fi

\bibitem[Deville and S{\"a}rndal(1992)]{deville1992calibration}
Jean-Claude Deville and Carl-Erik S{\"a}rndal.
\newblock Calibration estimators in survey sampling.
\newblock \emph{Journal of the American Statistical Association}, 87\penalty0
  (418):\penalty0 376--382, 1992.

\bibitem[S{\"a}rndal(2007)]{sarndal2007calibration}
Carl-Erik S{\"a}rndal.
\newblock The calibration approach in survey theory and practice.
\newblock \emph{Survey Methodology}, 33\penalty0 (2):\penalty0 99--119, 2007.

\bibitem[Breslow et~al.(2009)Breslow, Lumley, Ballantyne, Chambless, and
  Kulich]{breslow2009using}
Norman~E Breslow, Thomas Lumley, Christie~M Ballantyne, Lloyd~E Chambless, and
  Michal Kulich.
\newblock Using the whole cohort in the analysis of case-cohort data.
\newblock \emph{American Journal of Epidemiology}, 169\penalty0 (11):\penalty0
  1398--1405, 2009.

\bibitem[Robins et~al.(1994)Robins, Rotnitzky, and Zhao]{robins1994estimation}
James~M Robins, Andrea Rotnitzky, and Lue~Ping Zhao.
\newblock Estimation of regression coefficients when some regressors are not
  always observed.
\newblock \emph{Journal of the American Statistical Association}, 89\penalty0
  (427):\penalty0 846--866, 1994.

\bibitem[Firth and Bennett(1998)]{firth1998robust}
David Firth and KE~Bennett.
\newblock Robust models in probability sampling.
\newblock \emph{Journal of the Royal Statistical Society: Series B (Statistical
  Methodology)}, 60\penalty0 (1):\penalty0 3--21, 1998.

\bibitem[Lumley et~al.(2011)Lumley, Shaw, and Dai]{lumley2011connections}
Thomas Lumley, Pamela~A Shaw, and James~Y Dai.
\newblock Connections between survey calibration estimators and semiparametric
  models for incomplete data.
\newblock \emph{International Statistical Review}, 79\penalty0 (2):\penalty0
  200--220, 2011.

\bibitem[Rubin(1996)]{rubin1996multiple}
Donald~B Rubin.
\newblock Multiple imputation after 18+ years.
\newblock \emph{Journal of the American Statistical Association}, 91\penalty0
  (434):\penalty0 473--489, 1996.

\bibitem[Marti and Chavance(2011)]{marti2011multiple}
Helena Marti and Michel Chavance.
\newblock Multiple imputation analysis of case--cohort studies.
\newblock \emph{Statistics in Medicine}, 30\penalty0 (13):\penalty0 1595--1607,
  2011.

\bibitem[Keogh and White(2013)]{keogh2013using}
Ruth~H Keogh and Ian~R White.
\newblock Using full-cohort data in nested case--control and case--cohort
  studies by multiple imputation.
\newblock \emph{Statistics in Medicine}, 32\penalty0 (23):\penalty0 4021--4043,
  2013.

\bibitem[Jung et~al.(2016)Jung, Harel, and Kang]{jung2016fitting}
Jinhyouk Jung, Ofer Harel, and Sangwook Kang.
\newblock Fitting additive hazards models for case-cohort studies: {A} multiple
  imputation approach.
\newblock \emph{Statistics in Medicine}, 35\penalty0 (17):\penalty0 2975--2990,
  2016.

\bibitem[Seaman et~al.(2012)Seaman, White, Copas, and Li]{seaman2012combining}
Shaun~R Seaman, Ian~R White, Andrew~J Copas, and Leah Li.
\newblock Combining multiple imputation and inverse-probability weighting.
\newblock \emph{Biometrics}, 68\penalty0 (1):\penalty0 129--137, 2012.

\bibitem[Morris et~al.(2014)Morris, White, and Royston]{morris2014tuning}
Tim~P Morris, Ian~R White, and Patrick Royston.
\newblock Tuning multiple imputation by predictive mean matching and local
  residual draws.
\newblock \emph{BMC {M}edical {R}esearch {M}ethodology}, 14\penalty0
  (1):\penalty0 75, 2014.

\bibitem[Han(2016)]{han2016combining}
Peisong Han.
\newblock Combining inverse probability weighting and multiple imputation to
  improve robustness of estimation.
\newblock \emph{Scandinavian Journal of Statistics}, 43\penalty0 (1):\penalty0
  246--260, 2016.

\bibitem[{Lumley}(2017)]{lumley2017robustness}
T.~{Lumley}.
\newblock {Robustness of semiparametric efficiency in nearly-true models for
  two-phase samples}.
\newblock \emph{ArXiv e-prints}, July 2017.
\newblock arXiv: 1707.05924.

\bibitem[Carroll et~al.(2006)Carroll, Ruppert, Stefanski, and
  Crainiceanu]{carroll2006}
Raymond~J Carroll, David Ruppert, Leonard~A Stefanski, and Ciprian~M
  Crainiceanu.
\newblock \emph{Measurement Error in Nonlinear Models: A Modern Perspective}.
\newblock Chapman and Hall/CRC, Boca Raton, 2006.

\bibitem[Rivera and Lumley(2016)]{rivera2016using}
C~Rivera and T~Lumley.
\newblock Using the whole cohort in the analysis of countermatched samples.
\newblock \emph{Biometrics}, 72\penalty0 (2):\penalty0 382--391, 2016.

\bibitem[Prentice and Pyke(1979)]{prentice1979logistic}
Ross~L Prentice and Ronald Pyke.
\newblock Logistic disease incidence models and case-control studies.
\newblock \emph{Biometrika}, 66\penalty0 (3):\penalty0 403--411, 1979.

\bibitem[Wild and Jiang(2013)]{wild2013missreg3}
C.~Wild and Y.~Jiang.
\newblock \emph{\texttt{missreg3}: Software for a class of response selective
  and missing data problem}, 2013.
\newblock R package version under 3.00 (URL:
  \url{https://www.stat.auckland.ac.nz/~wild/software.html}).

\bibitem[Scott and Wild(2006)]{scott2006calculating}
A.~J. Scott and C.~J. Wild.
\newblock Calculating efficient semiparametric estimators for a broad class of
  missing-data problems.
\newblock In Eds~EP Liski, J~Isotalo, J~Niemel{\"a}, S~Puntanen, and G~P~H
  Styan, editors, \emph{Festschrift for Tarmo Pukkila on his 60th birthday},
  pages 301--314, 2006.

\bibitem[Cao-Abad(1991)]{cao1991rate}
R~Cao-Abad.
\newblock Rate of convergence for the wild bootstrap in nonparametric
  regression.
\newblock \emph{The Annals of Statistics}, 19\penalty0 (4):\penalty0
  2226--2231, 1991.

\bibitem[Mammen(1993)]{mammen1993bootstrap}
Enno Mammen.
\newblock Bootstrap and wild bootstrap for high dimensional linear models.
\newblock \emph{The Annals of Statistics}, 21\penalty0 (1):\penalty0 255--285,
  1993.

\bibitem[Hardle and Mammen(1993)]{hardle1993comparing}
Wolfgang Hardle and Enno Mammen.
\newblock Comparing nonparametric versus parametric regression fits.
\newblock \emph{The Annals of Statistics}, 21\penalty0 (4):\penalty0
  1926--1947, 1993.

\bibitem[Lumley(2011)]{lumley2011complex}
Thomas Lumley.
\newblock \emph{Complex Surveys: A Guide to Analysis using R}, volume 565.
\newblock John Wiley \& Sons, 2011.

\bibitem[Breslow and Chatterjee(1999)]{breslow1999design}
Norman~E Breslow and Nilanjan Chatterjee.
\newblock Design and analysis of two-phase studies with binary outcome applied
  to {Wilms tumour} prognosis.
\newblock \emph{Journal of the Royal Statistical Society: Series C (Applied
  Statistics)}, 48\penalty0 (4):\penalty0 457--468, 1999.

\bibitem[Freedman(2009)]{freedman2009}
David~A Freedman.
\newblock Diagnostics cannot have much power against general alternatives.
\newblock \emph{International Journal of Forecasting}, 25\penalty0
  (4):\penalty0 833--839, 2009.

\bibitem[Watson and Holmes(2016)]{watson2016}
James Watson and Chris Holmes.
\newblock Approximate models and robust decisions.
\newblock \emph{Statistical Science}, 31\penalty0 (4):\penalty0 465--489, 2016.

\bibitem[Box et~al.(2005)Box, Hunter, and Hunter]{box2005statistics}
George~EP Box, J~Stuart Hunter, and William~G Hunter.
\newblock \emph{Statistics for Experimenters}.
\newblock {Hoboken, NJ: Wiley}, 2005.

\bibitem[Stone(1985)]{stone1985additive}
Charles~J Stone.
\newblock Additive regression and other nonparametric models.
\newblock \emph{The Annals of Statistics}, 13\penalty0 (2):\penalty0 689--705,
  1985.

\bibitem[Freedman(2010)]{freedman2010statistical}
David~A Freedman.
\newblock \emph{Statistical Models and Causal Inference: A Dialogue with the
  Social Sciences}.
\newblock Cambridge University Press, 2010.

\bibitem[Hart(2013)]{hart2013nonparametric}
Jeffrey Hart.
\newblock \emph{Nonparametric Smoothing and Lack-of-fit Tests}.
\newblock Springer Science \& Business Media, 2013.

\bibitem[Li and Racine(2007)]{li2007nonparametric}
Qi~Li and Jeffrey~Scott Racine.
\newblock \emph{Nonparametric Econometrics: Theory and Practice}.
\newblock Princeton University Press, 2007.

\bibitem[McIsaac and Cook(2015)]{mcisaac2015adaptive}
Michael~A McIsaac and Richard~J Cook.
\newblock Adaptive sampling in two-phase designs: A biomarker study for
  progression in arthritis.
\newblock \emph{Statistics in Medicine}, 34\penalty0 (21):\penalty0 2899--2912,
  2015.

\bibitem[Racine and Hayfield(2018)]{racine2018np}
J.~S. Racine and T.~Hayfield.
\newblock \emph{np: Nonparametric kernel smoothing methods for mixed data
  types}, 2018.
\newblock R package version 0.60-9 (URL:
  \url{https://CRAN.R-project.org/package=np}).

\end{thebibliography}

\clearpage

\begin{table}[htbp]
    \caption{\label{table1} Relative performance of the maximum likelihood (MLE), design-based estimator (IPW), parametric imputation (MI-P) and bootstrap resampling (MI-B) imputation estimators  in the case-control design with cohort size $N=10^4$, case-control subset with $n=110$ in average, $M=100$ imputations, and $1000$ Monte Carlo runs. We report the root-mean squared error ($\sqrt{\textrm{MSE}}$) for $\beta=1$, its bias and variance decomposition \eqref{bias-var}, and the empirical power to reject the nearly-true model \eqref{nearly-true1} through the most powerful (MP) test and the goodness-of-fit test of linear fits.\citep{li2007nonparametric, hart2013nonparametric}}
    \centering
    \begin{tabular}{cc rrrr c cc}
	\hline
	\multirow{2}{*}{$(\beta_0, \delta_0)$}	&	\multirow{2}{*}{Criterion}	&	\multicolumn{4}{c}{Estimation performance}	&	&	\multicolumn{2}{c}{Empirical power$^\dagger$}\\
	\cline{3-6}\cline{8-9}
		&	&	{MLE}	&	{IPW} 	&	MI-P	&	MI-B	&	&	MP test	&	 Lin. test\\
	\hline
	\multirow{3}{*}{(1, 0)}		
		&	$\sqrt\textrm{MSE}$	&	0.145	&	0.239	&	0.140	&	0.240	&	&	\multirow{3}{*}{0.046}	&	\multirow{3}{*}{0.042}\\
		&	Bias				&	0.014	&	0.071	&	0.011	&	0.071	&	&	\\	
		& 	$\sqrt\textrm{Var}$	&	0.144	&	0.229	&	0.140	&	0.229	&	&	\\		
	\hline	
	\multirow{3}{*}{(0.844, 0.700)}
		&	$\sqrt\textrm{MSE}$	&	0.148	&	0.229	&	0.147	&	0.229	&	&	\multirow{3}{*}{0.202}	&	\multirow{3}{*}{0.042}\\
		&	Bias				&	-0.067	&	0.064	&	-0.077	&	0.064	&	&	\\	
		& 	$\sqrt\textrm{Var}$	&	0.132	&	0.219	&	0.125	&	0.219	&	&	\\			
	\hline
	\multirow{3}{*}{(0.692, 1.400)}
		&	$\sqrt\textrm{MSE}$	&	0.199	&	0.217	&	0.204	&	0.217	&	&	\multirow{3}{*}{0.410}	&	\multirow{3}{*}{0.061}\\
		&	Bias				&	-0.156	&	0.054	&	-0.168	&	0.054	&	&	\\	
		& 	$\sqrt\textrm{Var}$	&	0.124	&	0.211	&	0.116	&	0.211	&	&	\\			
	\hline
	\multirow{3}{*}{(0.541, 2.100)}
		&	$\sqrt\textrm{MSE}$	&	0.257	&	0.201	&	0.262	&	0.201	&	&	\multirow{3}{*}{0.683}	&	\multirow{3}{*}{0.156}\\
		&	Bias				&	-0.233	&	0.047	&	-0.242	&	0.047	&	&	\\	
		& 	$\sqrt\textrm{Var}$	&	0.109	&	0.196	&	0.102	&	0.195	&	&	\\			
	\hline
	\multirow{3}{*}{(0.381, 2.800)}
		&	$\sqrt\textrm{MSE}$	&	0.317	&	0.206	&	0.320	&	0.206	&	&	\multirow{3}{*}{0.905}	&	\multirow{3}{*}{0.382}\\
		&	Bias				&	-0.301	&	0.056	&	-0.306	&	0.056	&	&	\\	
		& 	$\sqrt\textrm{Var}$	&	0.098	&	0.199	&	0.093	&	0.199	&	&	\\			
	\hline
    \end{tabular}
    \begin{flushleft}
    	\item $^\dagger$$P_n$ and $Q_n$ are likelihood functions at $\theta_0 = (\alpha_0, \beta_0, \delta_0)$ and $\theta^* = (\alpha, \beta)$, respectively.
    \end{flushleft}
\end{table}

\begin{table}[htbp]
    \caption{\label{table2} Multiple imputation in two-stage analysis with continuous surrogates when $Z = X + \varepsilon$ for independent $\varepsilon \sim N(0,1)$. We compare relative performance of the maximum likelihood (MLE), standard raking, regression calibration (RC), multiple imputations (MI) using either the wild bootstrap or Bayesian approach, and the proposed multiple imputation with raking (MIR) estimators for a two-phase design with cohort size $N=5000$, phase 2 subset $|S_2|=750$ in average, $M=100$ imputations, and $1000$ Monte Carlo runs. We report the root-mean squared error ($\sqrt{\textrm{MSE}}$) for $\beta=1$, its bias and variance decomposition \eqref{bias-var}, and the empirical power to reject the nearly-true model \eqref{nearly-true2} through the most powerful (MP) test and the goodness-of-fit test of linear fits.\citep{hart2013nonparametric, li2007nonparametric}}  
    \centering
    \resizebox{\columnwidth}{!}{%
    \begin{tabular}{cc cccccccc ccc}
	\hline
	\multirow{3}{*}{$(\beta_0, \delta_0)$}	&	\multirow{3}{*}{Criterion}	&	\multicolumn{8}{c}{Estimation performance}	&	\multirow{3}{*}{\multirow{1}{*}{Abs Corr$^\dagger$}}	&	\multicolumn{2}{c}{\multirow{1}{*}{Empirical power$^\ddagger$}}\\
	\cline{3-10}\cline{12-13}
		&	&	\multirow{2}{*}{MLE}	&	\multirow{2}{*}{Raking}	&	\multirow{2}{*}{RC} 	&	\multicolumn{2}{c}{MI}	& &	\multicolumn{2}{c}{MIR}	&	&	\multirow{2}{*}{MP test}	&	\multirow{2}{*}{Lin. test}\\
	\cline{6-7}\cline{9-10}
		&	&	&	&	&	Boot	&	Bayes	& &	Boot	&	Bayes\\
	\hline
	\multirow{3}{*}{(1, 0)}		
		&	$\sqrt\textrm{MSE}$	&	0.019	&	0.038	&	0.017	&	0.019	&	0.019	& &	0.034	&	0.034	&	\multirow{3}{*}{-}	&	\multirow{3}{*}{0.052}	&	\multirow{3}{*}{0.065}\\
		&	Bias				&	0.004	&	0.000	&	0.000	&	0.002	&	-0.003	& &	0.001	&	0.001	&	&	\\	
		& 	$\sqrt\textrm{Var}$	&	0.019	&	0.038	&	0.017	&	0.018	&	0.018	& &	0.034	&	0.034	&	&	\\				
	\hline
	\multirow{3}{*}{(0.951, 0.068)}		
		&	$\sqrt\textrm{MSE}$	&	0.033	&	0.037	&	0.022	&	0.023	&	0.026	& &	0.033	&	0.033	&	\multirow{3}{*}{0.480}	&	\multirow{3}{*}{0.140}	&	\multirow{3}{*}{0.078}\\
		&	Bias				&	-0.027	&	0.000	&	-0.014	&	-0.014	&	-0.019	& &	0.001	&	0.001	&	&	\\	
		& 	$\sqrt\textrm{Var}$	&	0.018	&	0.037	&	0.017	&	0.018	&	0.018	& &	0.033	&	0.033	&	&	\\				
	\hline
	\multirow{3}{*}{(0.904, 0.131)}		
		&	$\sqrt\textrm{MSE}$	&	0.058	&	0.036	&	0.032	&	0.034	&	0.039	& &	0.033	&	0.033	&	\multirow{3}{*}{0.496}	&	\multirow{3}{*}{0.407}	&	\multirow{3}{*}{0.089}\\
		&	Bias				&	-0.056	&	0.000	&	-0.027	&	-0.029	&	-0.034	& &	0.001	&	0.001	&	&	\\	
		& 	$\sqrt\textrm{Var}$	&	0.018	&	0.036	&	0.017	&	0.018	&	0.018	& &	0.033	&	0.033	&	&	\\				
	\hline
	\multirow{3}{*}{(0.861, 0.191)}		
		&	$\sqrt\textrm{MSE}$	&	0.084	&	0.036	&	0.042	&	0.047	&	0.052	& &	0.032	&	0.032	&	\multirow{3}{*}{0.497}	&	\multirow{3}{*}{0.698}	&	\multirow{3}{*}{0.108}\\
		&	Bias				&	-0.082	&	-0.001	&	-0.038	&	-0.043	&	-0.048	& &	0.001	&	0.001	&	&	\\	
		& 	$\sqrt\textrm{Var}$	&	0.018	&	0.036	&	0.017	&	0.018	&	0.018	& &	0.032	&	0.032	&	&	\\				
	\hline
	\multirow{3}{*}{(0.820, 0.247)}		
		&	$\sqrt\textrm{MSE}$	&	0.108	&	0.035	&	0.052	&	0.059	&	0.064	& &	0.032	&	0.032	&	\multirow{3}{*}{0.496}	&	\multirow{3}{*}{0.893}	&	\multirow{3}{*}{0.142}\\
		&	Bias				&	-0.107	&	0.000	&	-0.049	&	-0.057	&	-0.062	& &	0.001	&	0.001	&	&	\\	
		& 	$\sqrt\textrm{Var}$	&	0.017	&	0.035	&	0.017	&	0.018	&	0.018	& &	0.032	&	0.032	&	&	\\				
	\hline
	\multirow{3}{*}{(0.781, 0.3)}		
		&	$\sqrt\textrm{MSE}$	&	0.132	&	0.035	&	0.062	&	0.072	&	0.077	& &	0.032	&	0.032	&	\multirow{3}{*}{0.495}	&	\multirow{3}{*}{0.978}	&	\multirow{3}{*}{0.189}\\
		&	Bias				&	-0.131	&	-0.001	&	-0.060	&	-0.069	&	-0.074	& &	0.001	&	0.001	&	&	\\	
		& 	$\sqrt\textrm{Var}$	&	0.017	&	0.035	&	0.017	&	0.018	&	0.018	& &	0.032	&	0.032	&	&	\\				
	\hline
    \end{tabular}
    }
    \begin{flushleft}
    	\item $^{\dagger,\ddagger}$The absolute value of the  correlation between $\hat{\beta}_{\textrm{MLE}} - \hat{\beta}_{\textrm{Raking}}$ and $\log {Q}_n- \log {P}_n$,
    	\item where $P_n$ and $Q_n$ are likelihood functions at $\theta_0 = (\alpha_0, \beta_0, \delta_0)$ and $\theta^* = (\alpha, \beta)$, respectively.
    \end{flushleft}
\end{table}

\begin{table}[htbp]
    \caption{\label{table3} Multiple imputation in two-stage analysis with continuous surrogates when $Z = \eta X$ for independent $\eta \sim \Gamma(4,4)$. We compare relative performance of the maximum likelihood (MLE), standard raking, regression calibration (RC), multiple imputations using (MI) either the wild bootstrap or Bayesian approach, and the proposed multiple imputation with raking (MIR) estimators for a two-phase design with cohort size $N=5000$, phase 2 subset $|S_2|=750$ in average, $M=100$ imputations, and $1000$ Monte Carlo runs. We report the root-mean squared error ($\sqrt{\textrm{MSE}}$) for $\beta=1$, its bias and variance decomposition \eqref{bias-var}, and the empirical power to reject the nearly-true model \eqref{nearly-true2} through the most powerful (MP) test and the goodness-of-fit test of linear fits.\citep{hart2013nonparametric, li2007nonparametric}}  
    \centering
    \resizebox{\columnwidth}{!}{%
    \begin{tabular}{cc cccccccc ccc}
	\hline
	\multirow{3}{*}{$(\beta_0, \delta_0)$}	&	\multirow{3}{*}{Criterion}	&	\multicolumn{8}{c}{Estimation performance}	&	\multirow{3}{*}{\multirow{1}{*}{Abs Corr$^\dagger$}}	&	\multicolumn{2}{c}{\multirow{1}{*}{Empirical power$^\ddagger$}}\\
	\cline{3-10}\cline{12-13}
		&	&	\multirow{2}{*}{MLE}	&	\multirow{2}{*}{Raking}	&	\multirow{2}{*}{RC} 	&	\multicolumn{2}{c}{MI}	& &	\multicolumn{2}{c}{MIR}	&	&	\multirow{2}{*}{MP test}	&	\multirow{2}{*}{Lin. test}\\
	\cline{6-7}\cline{9-10}
		&	&	&	&	&	Boot	&	Bayes	& &	Boot	&	Bayes\\
	\hline
	\multirow{3}{*}{(1, 0)}		
		&	$\sqrt\textrm{MSE}$	&	0.018	&	0.030	&	0.216	&	0.099	&	0.094	& &	0.029	&	0.029	&	\multirow{3}{*}{-}	&	\multirow{3}{*}{0.048}	&	\multirow{3}{*}{0.056}\\
		&	Bias				&	0.006	&	0.001	&	0.215	&	0.097	&	0.092	& &	0.002	&	0.002	&	&	\\	
		& 	$\sqrt\textrm{Var}$	&	0.017	&	0.030	&	0.013	&	0.018	&	0.018	& &	0.029	&	0.029	&	&	\\	
	\hline
	\multirow{3}{*}{(1.045, -0.068)}		
		&	$\sqrt\textrm{MSE}$	&	0.040	&	0.030	&	0.227	&	0.111	&	0.106	& &	0.029	&	0.029	&	\multirow{3}{*}{0.585}	&	\multirow{3}{*}{0.149}	&	\multirow{3}{*}{0.062}\\
		&	Bias				&	0.036	&	0.001	&	0.227	&	0.109	&	0.104	& &	0.002	&	0.002	&	&	\\	
		& 	$\sqrt\textrm{Var}$	&	0.018	&	0.030	&	0.013	&	0.018	&	0.018	& &	0.029	&	0.029	&	&	\\	
	\hline
	\multirow{3}{*}{(1.087, -0.131)}		
		&	$\sqrt\textrm{MSE}$	&	0.068	&	0.031	&	0.239	&	0.123	&	0.117	& &	0.030	&	0.030	&	\multirow{3}{*}{0.584}	&	\multirow{3}{*}{0.427}	&	\multirow{3}{*}{0.075}\\
		&	Bias				&	0.065	&	0.001	&	0.238	&	0.121	&	0.116	& &	0.002	&	0.002	&	&	\\	
		& 	$\sqrt\textrm{Var}$	&	0.018	&	0.031	&	0.013	&	0.018	&	0.018	& &	0.030	&	0.030	&	&	\\	
	\hline
	\multirow{3}{*}{(1.127, -0.191)}		
		&	$\sqrt\textrm{MSE}$	&	0.095	&	0.032	&	0.249	&	0.134	&	0.128	& &	0.031	&	0.031	&	\multirow{3}{*}{0.585}	&	\multirow{3}{*}{0.697}	&	\multirow{3}{*}{0.099}\\
		&	Bias				&	0.093	&	0.001	&	0.249	&	0.133	&	0.127	& &	0.002	&	0.002	&	&	\\	
		& 	$\sqrt\textrm{Var}$	&	0.018	&	0.032	&	0.014	&	0.018	&	0.018	& &	0.030	&	0.031	&	&	\\	
	\hline
	\multirow{3}{*}{(1.165, -0.247)}		
		&	$\sqrt\textrm{MSE}$	&	0.121	&	0.032	&	0.259	&	0.144	&	0.139	& &	0.031	&	0.031	&	\multirow{3}{*}{0.583}	&	\multirow{3}{*}{0.890}	&	\multirow{3}{*}{0.136}\\
		&	Bias				&	0.119	&	0.001	&	0.259	&	0.143	&	0.138	& &	0.002	&	0.002	&	&	\\	
		& 	$\sqrt\textrm{Var}$	&	0.019	&	0.032	&	0.014	&	0.019	&	0.019	& &	0.031	&	0.031	&	&	\\	
	\hline
	\multirow{3}{*}{(1.200, -0.3)}		
		&	$\sqrt\textrm{MSE}$	&	0.146	&	0.033	&	0.269	&	0.155	&	0.149	& &	0.032	&	0.032	&	\multirow{3}{*}{0.580}	&	\multirow{3}{*}{0.967}	&	\multirow{3}{*}{0.179}\\
		&	Bias				&	0.145	&	0.001	&	0.268	&	0.154	&	0.148	& &	0.003	&	0.002	&	&	\\	
		& 	$\sqrt\textrm{Var}$	&	0.019	&	0.033	&	0.014	&	0.019	&	0.019	& &	0.032	&	0.032	&	&	\\	
	\hline
    \end{tabular}
    }
    \begin{flushleft}
    	\item $^{\dagger,\ddagger}$The absolute value of the  correlation between $\hat{\beta}_{\textrm{MLE}} - \hat{\beta}_{\textrm{Raking}}$ and $\log {Q}_n- \log {P}_n$, 
        \item where $P_n$ and $Q_n$ are likelihood functions at $\theta_0 = (\alpha_0, \beta_0, \delta_0)$ and $\theta^* = (\alpha, \beta)$, respectively.
    \end{flushleft}
\end{table}

\begin{table}[htbp]
    \caption{\label{table4} The National Wilms Tumor Study data example. We compare relative performance of the maximum likelihood (MLE), standard raking, multiple imputation (MI) using the wild bootstrap (MI), and the proposed multiple imputation with raking (MIR) estimators for a two-phase design with cohort size $N=3915$, phase 2 subset $|S_2|=1338$, $M=100$ imputations, and $1000$ Monte Carlo runs. We report the root-mean squared error ($\sqrt{\textrm{MSE}}$) for the parameter estimate obtained from the full cohort analysis of the outcome model \eqref{wilms-model}, and its bias and variance decomposition \eqref{bias-var}.}
    \centering
    \begin{tabular}{cc ccccc c}
	\hline
	\multirow{2}{*}{Method}	&	\multirow{2}{*}{Criterion}	&	\multicolumn{5}{c}{Estimation performance by regressor}	&	\multirow{1}{*}{Sum of }\\
	\cline{3-7}
		&	&	Hstg$^1$	&	Stage$^2$	&	Age$^3$	&	Diam$^4$	&	H$\ast$S$^5$	&	Squares\\
	\hline	
	\multirow{3}{*}{MLE}	
		&	$\sqrt{\textrm{MSE}}$	&	1.768	&	0.777	&	0.014	&	0.014	&	0.605	&	4.096\\
		&	Bias					&	-1.768	&	-0.777	&	-0.007	&	-0.012	&	0.603	&	4.091\\
		&	$\sqrt{\textrm{Var}}$		&	0.031	&	0.023	&	0.013	&	0.008	&	0.051	&	0.005\\
	\hline
	\multirow{3}{*}{Raking}	
		&	$\sqrt{\textrm{MSE}}$	&	0.129	&	0.022	&	0.006	&	0.003	&	0.203	&	0.059\\
		&	Bias					&	0.021	&	-0.002	&	0.000	&	0.001	&	-0.050	&	0.003\\
		&	$\sqrt{\textrm{Var}}$		&	0.127	&	0.022	&	0.006	&	0.003	&	0.197	&	0.056\\
	\hline
	\multirow{3}{*}{MI}	
		&	$\sqrt{\textrm{MSE}}$	&	0.146	&	0.015	&	0.003	&	0.002	&	0.175	&	0.052\\
		&	Bias					&	0.055	&	-0.004	&	0.003	&	0.002	&	-0.042	&	0.005\\
		&	$\sqrt{\textrm{Var}}$		&	0.135	&	0.015	&	0.002	&	0.001	&	0.170	&	0.047\\
	\hline
	\multirow{3}{*}{MIR}	
		&	$\sqrt{\textrm{MSE}}$	&	0.124	&	0.022	&	0.006	&	0.003	&	0.189	&	0.051\\
		&	Bias					&	0.018	&	0.006	&	0.001	&	0.001	&	-0.038	&	0.002\\
		&	$\sqrt{\textrm{Var}}$		&	0.122	&	0.021	&	0.006	&	0.003	&	0.185	&	0.049\\
	\hline
	\multirow{2}{*}{Full cohort}
		&	Estimate		&	1.193	&	0.285	&	0.089	&	0.028	&	0.816	&	\multicolumn{1}{c}{-}\\
		&	Std. Error		&	0.156	&	0.105	&	0.017	&	0.012	&	0.227	&	\multicolumn{1}{c}{-}\\
	\hline
    \end{tabular}
    \begin{flushleft}
    	\item \hspace{1in}$^1$Unfavorable histology versus favorable; $^2$Disease stage III/IV versus I/II; 
    	\item \hspace{1in}$^3$Year at diagnosis; $^4$Tumor diameter (cm); $^5$Histology$\ast$Stage.
    \end{flushleft}
\end{table}

\begin{figure}[htbp]
        \centering
        \centerline{
        \includegraphics[width=0.7\textwidth]{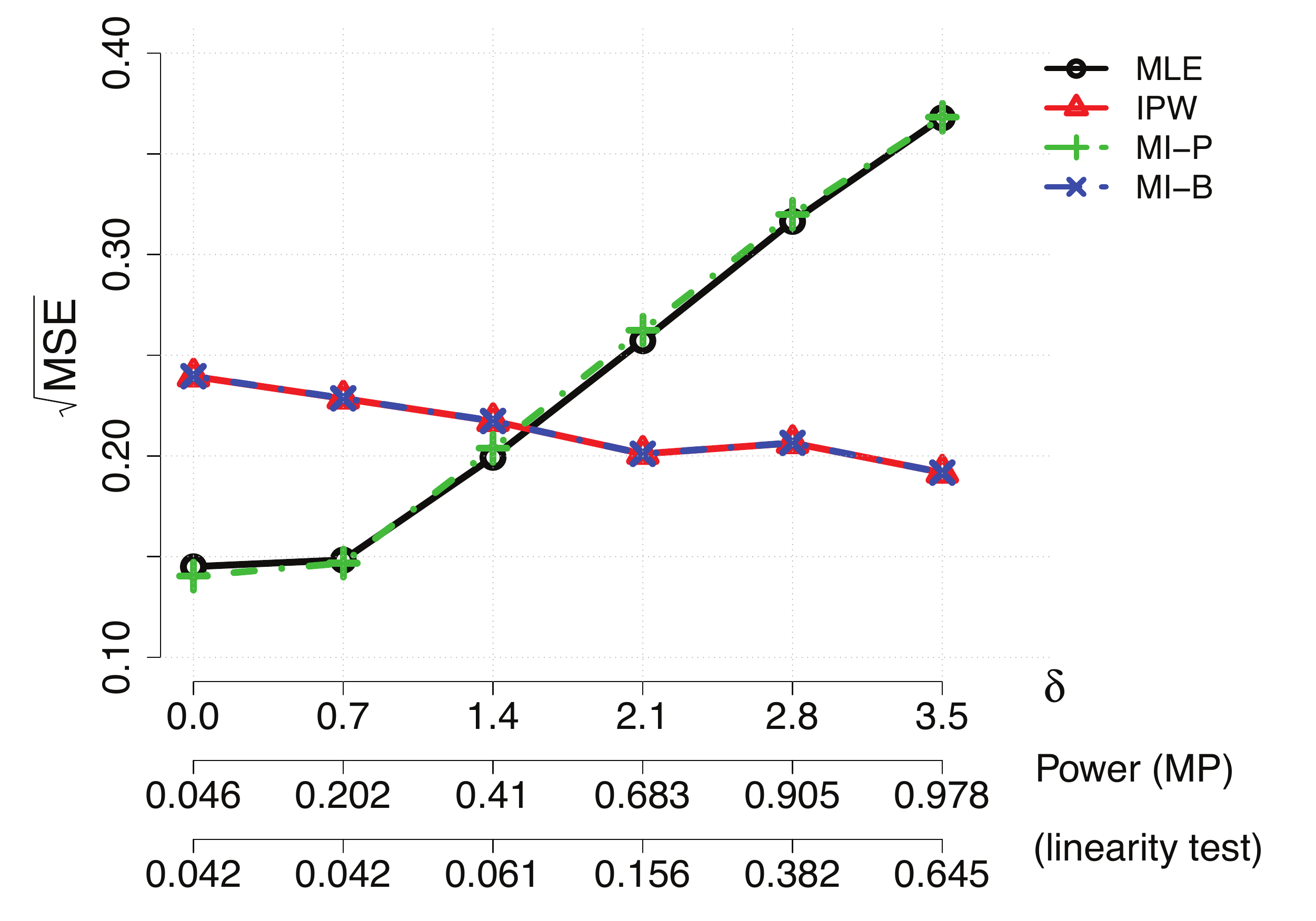}
        }
	\caption{Illustration of Table \ref{table1}. Relative performance of the maximum likelihood (MLE), design-based  estimator (IPW), parametric  imputation  (MI-P)  and  bootstrap  resampling (MI-B) imputation estimators  in the case-control design.}
	\label{figure1}
\end{figure}

\begin{figure}[htbp]
        \centering
        \centerline{
        \includegraphics[width=0.7\textwidth]{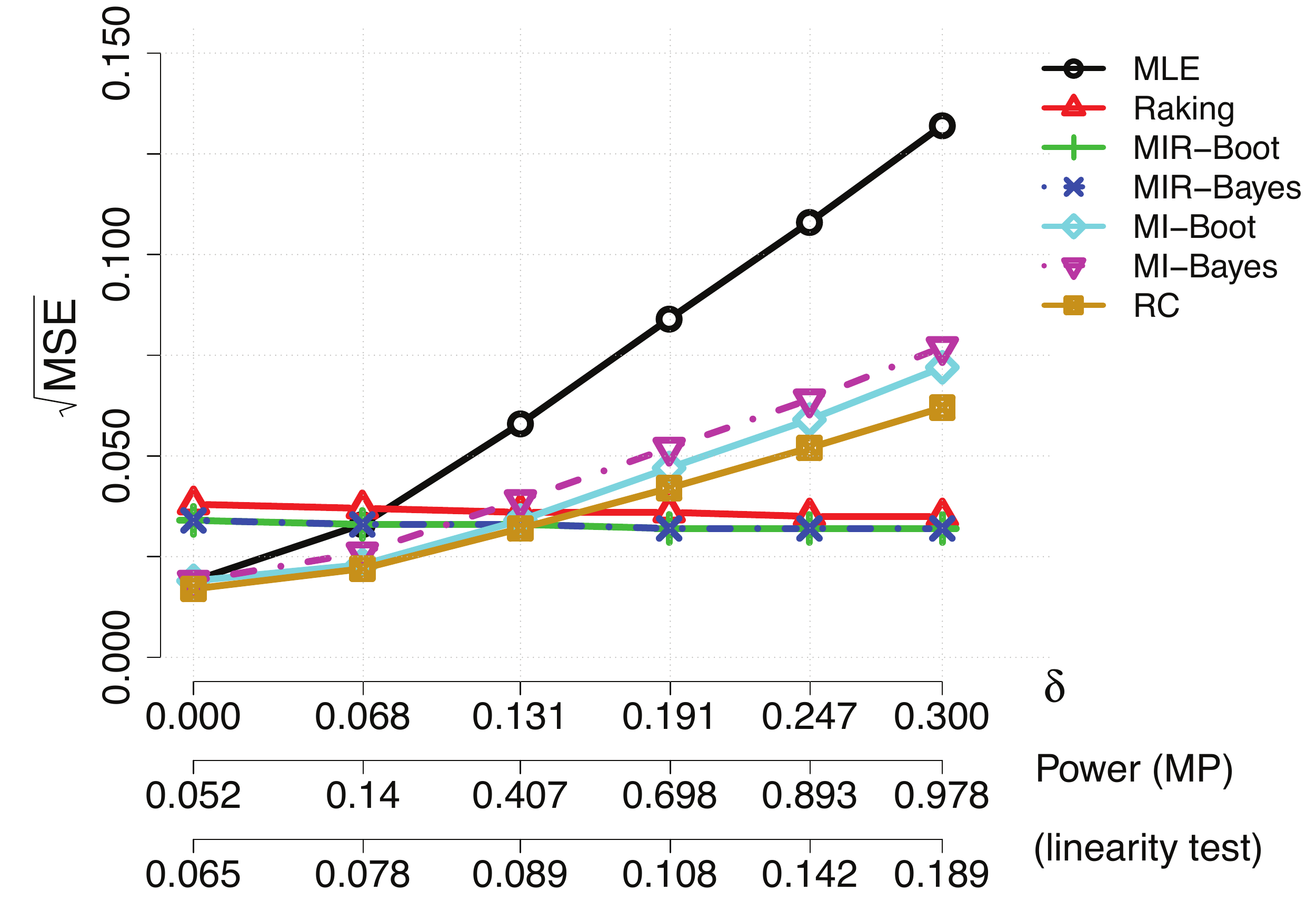}
        }
	\caption{Illustration of Table \ref{table2}. Relative performance of the maximum likelihood (MLE), standard raking, regression calibration (RC), multiple imputations (MI) using either the wild bootstrap or Bayesian approach, and the proposed multiple imputation with raking (MIR) estimators in two-stage analysis with continuous surrogates when $Z = X+\varepsilon$ for independent $\varepsilon \sim N(0,1)$.}
	\label{figure2}
\end{figure}

\begin{figure}[htbp]
        \centering
        \centerline{
        \includegraphics[width=0.7\textwidth]{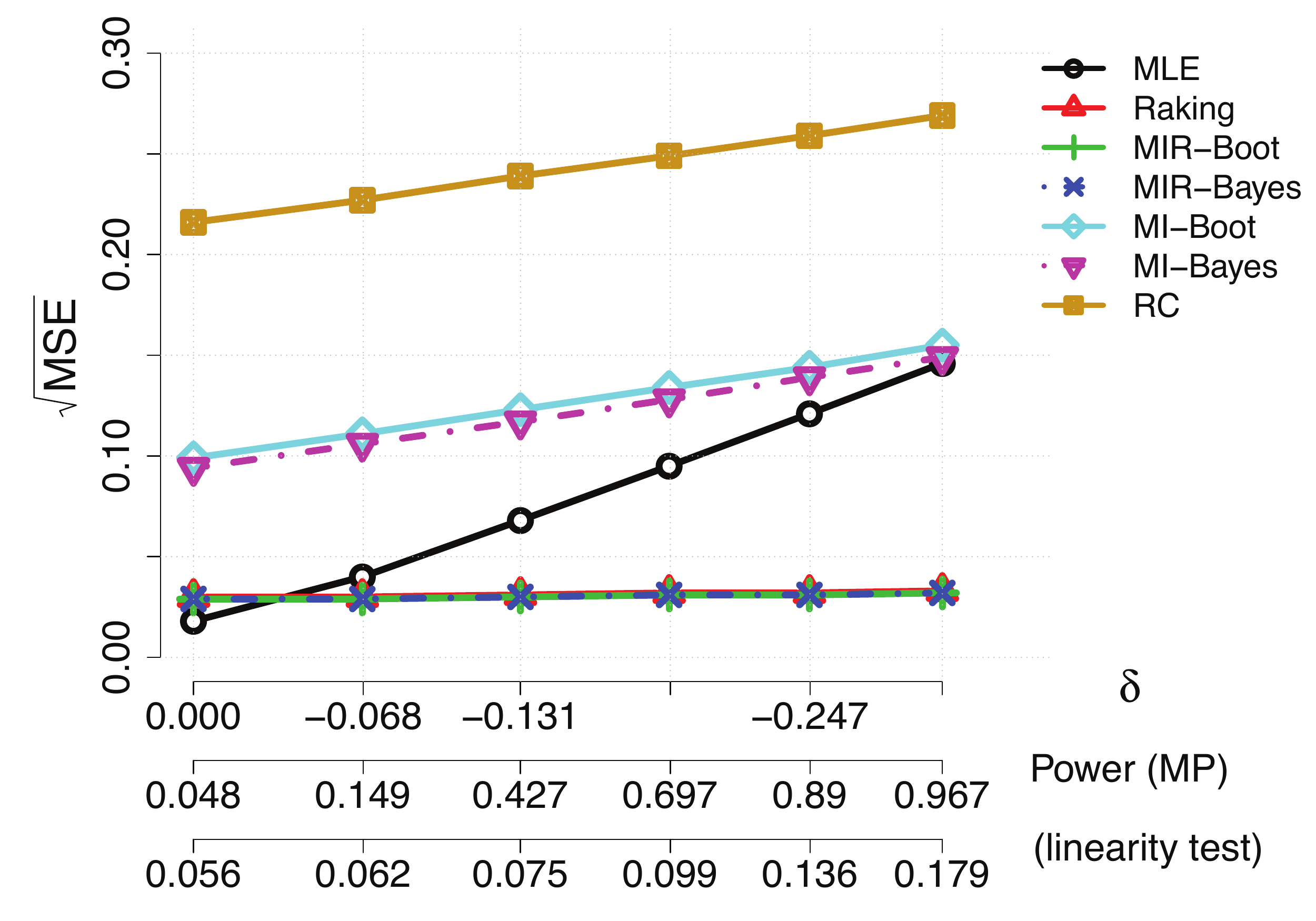}
        }
	\caption{Illustration of Table \ref{table3}. Relative performance of the maximum likelihood (MLE), standard raking, regression calibration (RC), multiple imputations (MI) using either the wild bootstrap or Bayesian approach, and the proposed multiple imputation with raking (MIR) estimators in two-stage analysis with continuous surrogates when $Z = \eta X$ for independent $\eta \sim \Gamma(4,4)$.}
	\label{figure3}
\end{figure}

\clearpage

\appendix

\section{Details of implementation} \label{appendix}

\subsection{Imputation}
\label{App-cal}

The wild bootstrap multiple imputation estimator is computed as follows:
\begin{itemize}
	\item[W1.] Generate ${X}_i^\ast = \hat{X}_i + V_i \hat{e}_i$ for $i \in S_2$, where $\hat{e}_i$ are residuals from (R2) and $V_i$ is an independent dichotomous random variable that takes on the value $(1+\sqrt{5})/2$ with probability $(\sqrt{5}-1)/ (2\sqrt{5})$, otherwise $(1-\sqrt{5})/2$, so that $\mathbb{E}V=0$ and $\textrm{Var}(V) = 1$.
	\item[W2.] Find an imputation model regressing ${X}_i^\ast$ on $Y_i$ and $Z_i$ for $i \in S_2$.
	\item[W3.] Resample $\hat{X}_i^\ast \sim N\big({\nu}(Y_i, Z_i), {\tau}^2(Y_i, Z_i)\big)$ independently for $i \in S_1$, where the mean and variance functions $\nu(Y_i, Z_i)\equiv\mathbb{E}(X | Y=y,Z=z)$ and ${\tau}^2(Y_i, Z_i)\equiv\textrm{Var}(X | Y=y,Z=z)$ are estimated from the model in (W2).
	\item[W4.] Fit the nearly-true model \eqref{nearly-true2} using $\{ (Y_i, \hat{X}_i^\ast) :  1 \leq i \leq N\}$, where $\hat{X}_i^\ast = X_i$ for $i \in S_2$.
	\item[W5.] Repeat (W1)--(W4) and take the average of multiple estimates of parameters.
\end{itemize}
We employ a parametric Bayesian resampling technique as follows:
\begin{itemize}
	\item[B1.] Find a posterior distribution of parameters $(a,b,c,\tau^2)$ for the imputation model used in (R1) given the second phase sample $\mathcal{X}_{II}$.
	\item[B2.] Generate $(a^\ast,b^\ast,c^\ast,\tau_\ast^2)$ from the posterior distribution in (B1).
	\item[B3.] Resample ${X}_i^\ast \sim N\big(a^\ast + b^\ast Y_i + c^\ast Z_i, {\tau}_\ast^2\big)$ independently for $i \in S_1$.
	\item[B4.] Fit the nearly-true model \eqref{nearly-true2} using $\{ (Y_i, \hat{X}_i^\ast) :  1 \leq i \leq N\}$, where $\hat{X}_i^\ast = X_i$ for $i \in S_2$.
	\item[B5.] Repeat (B1)--(B4) and take the average of multiple estimates of parameters.
\end{itemize}
For the prior distribution of $(a,b,c,\tau^2)$, we adopt a non-informative prior $p(a,b,c,\tau^2) \propto 1/\tau^2$. In (B2), we first generate $\tau_\ast^2 | \mathcal{X}_{II} \sim \Gamma^{-1}(a_n/2, b_n/2)$, where $a_n = |S_2| - 3$ and $b_n$ is the residual sum of squares from the linear regression model. Then, we generate $(a^\ast,b^\ast,c^\ast)^\top|\tau_\ast^2, \mathcal{X}_{II} \sim N_3\big( (\hat{a}, \hat{b}, \hat{c})^\top, \tau_\ast^2 (\Xi^\top \Xi)^{-1}\big)$, where $\Xi$ is the design matrix of the linear regression model in (R1) and $(\hat{a}, \hat{b}, \hat{c})$ is the corresponding estimate of the regression coefficient.

\subsection{Goodness-of-fit test}
\label{App-gof}

We use the wild bootstrap \citep{cao1991rate, mammen1993bootstrap,hardle1993comparing} together with kernel smoothing techniques in testing model specification of the parametric model. Suppose the true model is given by 
\begin{align}
    Y = m(X;\theta) + \varepsilon, \label{lof-eq1}
\end{align}
where $m$ is a known function depending of the parameter $\theta$ and $\varepsilon$ is a noise uncorrelated to $X$, that is $\mathbb{E}(\varepsilon | X) = 0$. In our study, we are mainly interested in in testing the null hypothesis such that
\begin{align}
    H_0 : m(X;\theta) = \alpha + \beta X \quad (a.e.) \nonumber
\end{align}
for some $\theta =(\alpha, \beta)^\top \in \mathbf{R}^2$. 
We note that under the null hypothesis $H_0$, estimation of $\mathbb{E}(Y|X=\cdot)$ in a fully nonparametric way regressing i.i.d. observations $Y_i$ on $X_i$, $1 \leq i \leq n$, is less efficient than we directly fit the parametric model \eqref{lof-eq1} based on the same sample. However, fitting the parametric model may suffers from inevitable bias when the model is misspecified as the sample size is increasing.\cite{hart2013nonparametric,li2007nonparametric}

From the above observation, we may test if the mean squared error quantifying the goodness-of-fit of the specified model \eqref{lof-eq1} is small compared to the nonparametric fits. Specifically, we measure  $\ell_n = \textrm{MSE}(\hat{\theta}) - \textrm{MSE}(\hat{m})$ and examine if the observed quantity $\ell_n$ is significantly small, where $\hat{m}(\cdot)$ is a univariate kernel regression estimator of $\mathbb{E}(Y|X=\cdot)$. Here, we choose the bandwidth for kernel smoothing based on leave-one-out cross validation criterion which empirically optimizes prediction performance of the kernel smoothed estimates and it can be easily implemented by using the \texttt{npregbw} function of the \texttt{np} package in R \citep{racine2018np}. Similarly to the previous ideas of the bootstrap resampling, the p-value of testing the null hypothesis $H_0$ is computed as below:
\begin{itemize}
	\item[T1.] Generate ${Y}_i^\ast = \hat{\alpha} + \hat{\beta}X_i + V_i \hat{e}_i$, $1 \leq i \leq n$, where $\hat{e}_i = Y_i - \hat{\alpha} + \hat{\beta}X_i$ and $V_i$ are random copies of an independent random variable $V$ which takes binary values by $(1+\sqrt{5})/2$ with probability $(\sqrt{5}-1)/ (2\sqrt{5})$, otherwise $(1-\sqrt{5})/2$ so that $\mathbb{E}V=0$ and $\textrm{Var}(V) = 1$.
	\item[T2.] Fit the parametric model with $(Y_1^\ast,X_1), \ldots, (Y_n^\ast, X_n)$ and let $\hat{\theta}^\ast = (\hat{\alpha}^\ast, \hat{\beta}^\ast)^\top$ be the resulting estimate of the parameter $\theta$. Compute the mean squared error $\textrm{MSE}(\hat{\theta}^\ast) = n^{-1}\sum_{i=i}^n (Y_i^\ast - \hat{\alpha}^\ast - \hat{\beta}^\ast X_i)^2$.
	\item[T3.] Find kernel smoothed fits $\hat{Y}^\ast = \hat{m}^\ast(X_i)$, $1 \leq i \leq n$, and compute the mean squared error $\textrm{MSE}(\hat{m}^\ast) = n^{-1}\sum_{i=i}^n (Y_i^\ast - \hat{m}^\ast(X_i))^2$.
	\item[T4.] Repeat (L1)--(L3) independently to obtain $\ell_n^\ast = \textrm{MSE}(\hat{\theta}^\ast) - \textrm{MSE}(\hat{m}^\ast)$ in multiple times to get an emirical distribution of $\ell_n$.
	\item[T5.] Compute the empirical p-value as the fraction of events $\ell_n^\ast > \ell_n$ occurred among repeated runs in (L4).
\end{itemize}

\end{document}